\documentclass{aa}
\usepackage{graphicx}
\newcommand{\lsim}{\raisebox{-5pt}{$\;\stackrel{\textstyle <}{\sim}\;$}}
\newcommand{\gsim}{\raisebox{-5pt}{$\;\stackrel{\textstyle >}{\sim}\;$}}

\begin{document}
   \title{\ion{Ar}{i}  as a tracer of  ionization evolution\thanks{Based on 
observations made with the ESO 8.2 m Kueyen 
telescope operated on Paranal Observatory, (Chile). ESO programmes 65.O-0299, 
66.A-0594(A), and 69.A-0299} 
}

   \author{G. Vladilo, M. Centuri\'on, V. D'Odorico, C. P\'eroux
          %\inst{1}
%          \and
%          B. Author2\inst{2} 
          }

   \offprints{G. Vladilo}

   \institute{
              Istituto Nazionale di Astrofisica, Osservatorio Astronomico di Trieste\\
              \email{vladilo@ts.astro.it}
                }

   \date{}

\titlerunning{\ion{Ar}{i}  as a tracer of  ionization evolution}

\authorrunning{Vladilo et al.}
 
 \abstract{
We present a study of Ar abundances in 15 damped Ly $\alpha$ systems (DLAs)
in the redshift interval $2.3 \leq z  \leq 3.4$. 
The sample includes 4 UVES/VLT measurements of \ion{Ar}{i} 
column densities presented here for the first time, together with 6 measurements
and 5 upper/lower limits collected from the literature  
(UVES/VLT and HIRES/Keck  observations).
The majority of DLAs  show significant underabundances
of Ar relative to other $\alpha$-capture elements with common nucleosynthetic origin. 
We show that neither dust depletion nor 
intervening \ion{H}{ii} regions inside DLAs
offer a viable justification to these underabundances. 
A natural explanation is found in the framework of
photoionization   models of   \ion{H}{i} regions embedded in
an ionizing continuum with varying spectral distribution.
At $z \sim 2.5$ the observed Ar deficiencies are large, 
[Ar/$\alpha$] $\simeq -0.6/-0.8$ dex, suggestive
of a hard, QSO-dominated  spectrum.
At $z \gsim 3$ the deficiencies
are instead small, suggestive of a soft, stellar-type spectrum,
though more data 
are needed to generalize this high-$z$ result.  
Should the change  of   Ar abundances with redshift
be governed by the
evolution of the UV stellar emission internal to DLAs,  
a   synchronization of the
star formation  in DLAs would be required, with a strong stellar
emission   at $z > 3$, but weak at $z < 3$.
This variation seems inconsistent  with the weak signal of evolution
indicated by abundance 
studies of DLAs.
More likely, the change  of Ar abundances is induced by the evolution   
of the UV metagalactic continuum, in which case 
the UV emission internal to DLAs must be small
(i.e. DLAs should have modest star formation rates) 
%the \ion{H}{i} regions in DLAs must be directly exposed to external radiation, 
and the external background  must become softer at $z > 3$.
%(e.g. dominated by Ly-break galaxies). 
The former requirement is   consistent with 
the modest evolution of DLAs abundances and 
the lack of Ly\,$\alpha$ and H\,$\alpha$ emissions associated with DLAs.
The latter requirement is consistent with
the observed evolution of \ion{Si}{iv}/\ion{C}{iv} ratios in the IGM,
the claims of high escape fraction of UV photons from Ly-break galaxies at $z \gsim 3$,
and the recent finding that the \ion{He}{ii} re-ionization seems to occur
between $z\sim 3.4$ and $z\sim 3$.
Comparison with results from local interstellar studies indicates
that Ar abundances can be used to trace the evolution of the
ionization history of the universe down to $z=0$, where 
[Ar/$\alpha$] $\sim -0.2$ dex.
We predict a rise of Ar abundances in the redshift range from
$z \simeq 2.3$ to $z=0$,
at the epoch at which the metagalactic field of galaxies overcomes
that of quasars. 
%Such rise   could be detected by a space-born UV telescope with large collecting area.
%
   \keywords{Diffuse radiation -- Intergalactic medium -- Ultraviolet: galaxies
   -- Cosmology: observations -- Quasars: absorption lines -- Quasars:
   individual (PKS 0528--250, Q0841+129, Q2343+1232)}
   }

   \maketitle
%
%________________________________________________________________

%%%%%%%%

\begin{center} 
\begin{table*} 
\scriptsize{
\caption{  Column densities$^a$ and relative abundances of  argon in Damped Ly $\alpha$ systems. }
\begin{tabular}{lllllllll}
\hline \hline
QSO & $z_\mathrm{abs}$ & log $N$(\ion{H}{i}) & log $N$(\ion{Ar}{i}) &
~~~[Ar/O] &  ~~~[Ar/S] &~~~[Ar/Si] & $\Delta_\mathrm{[Ar/Si]}^b$ & References \\  
\hline 
0528-2505	&	2.812	&	21.11$\pm$0.04	& 14.25$\pm$0.01	&		&$-$0.55$\pm$0.02		&	$-$0.63$\pm$0.04	& $-0.04$  	&	This work; P\'eroux et al. (2003)	\\ 
0841+129	&	2.375	& 21.00$\pm$0.10	&	13.46$\pm$0.10	&			& $-$0.70$\pm$0.10&$-$0.74$\pm$0.10	&	$-0.00$ 	&	This work; Centuri\'on et al.(2003)	\\ 
0841+129	&	2.476	&	20.79$\pm$0.10	& 13.34$\pm$0.12	&				&	$-$0.57$\pm$0.12	&$-$0.55$\pm$0.14	&	$-0.01$ & 	This work; Centuri\'on et al.(2003)	\\ 
2343+1232	&	2.431	&	20.18$\pm$0.05$^c$	& 13.22$\pm$0.05$^c$	&	 	&$-$0.78$\pm$0.05$^c$		&	$-$0.84$\pm$0.05$^c$	&  ~~~---	&	This work; D'Odorico et al.(2002) \\ 
\hline
0000-263	&	3.390	&	21.41$\pm$0.08	&	14.02$\pm$0.03	&	$-$0.03$\pm$0.03	& +0.00$\pm$0.04&	+0.00$\pm$0.04	& 	$-0.00$ 	&	Molaro et
al.(2000,2001)	\\ 
0201+365	&	2.462	&	20.38$\pm$0.05	& 14.08$\pm$0.03	&			& $-$0.53$\pm$0.03	&	$-$0.42$\pm$0.03	&	$-0.20$ 	&	Prochaska et
al.(2002a)	\\ 
0347-383	&	3.025	&	20.63$\pm$0.01	&	13.99$\pm$0.02	&	$-$0.43$\pm$0.02 &	& $-$0.21$\pm$0.02	&	$-0.00$  			&	Levshakov et al.(2002)	\\ 
0741+4741	& 3.017	&	20.48$\pm$0.10	&	13.13$\pm$0.02	&		&		&	$-$0.18$\pm$0.02	&	~~~--- 		&	Prochaska et al.(2002a)	\\ 
1759+7539	& 2.626	&	20.61$\pm$0.01$^c$	&	13.68$\pm$0.03$^c$	&		&	$-$0.71$\pm$0.13$^c$	&$-$0.68$\pm$0.04$^c$	& ~~~--- &
 Prochaska et al.(2002b)	\\  
2243$-$6031	&	2.330	&	20.24$\pm$0.02$^c$	& 13.35$\pm$0.05$^c$ &			 & $-$0.59$\pm$0.04$^c$	&$-$0.54$\pm$0.04$^c$	& $-0.00$ &
	L\'opez et al.(2002)	\\
\hline												
0001	&	3.000	&	20.70$\pm$0.05	&	$~~~<$13.38			&	$~~~<-$0.16	&		&	$~~~<-$0.03	&		 			&	Prochaska et al.(2002a)	\\
0336-0142	&	3.062	&	21.20$\pm$0.10	&	$~~~>$13.94			&				&	 	$~~~>-$0.37	& &					&	Prochaska et al.(2002a)	\\
0930+2858	&	3.235	&	20.30$\pm$0.10	&	$~~~<$12.96			&		&		&	$~~~<$+0.11	&		 				&	Prochaska et al.(2002a)	\\
1426+6039	&	2.827	&	20.30$\pm$0.04	&	$~~~<$13.43			&		&		&	$~~~<-$0.36	&	 				&	Prochaska et al.(2002a)	\\
2344+12	&	2.538	&	20.36$\pm$0.10	&	$~~~<$13.26			&			&	&	$~~~<$+0.12	&	 			&	Prochaska et al.(2002a)	\\
\hline
\end{tabular} 
\\ 
$^a$ Column densities are expressed in atoms cm$^{-2}$.
\\
$^b$ Additive dust correction term  for [Ar/Si] estimated using the set of
parameters S11(Vladilo 2002) and assuming zero dust depletion for Ar.
\\
$^c$  
Sum of  velocity components observed in  \ion{Ar}{i};
the fraction of $N$(\ion{Si}{ii}) measured in such components
over total $N$(\ion{Si}{ii}) has been used to scale down
the total $N(\ion{H}{i})$;  
components 6 and 7 in Prochaska et al. (2002b) have been considered
for the DLA in Q1759+7539;
components 3, 4 and 11 in L\'opez et al. (2002) 
for the DLA in Q2243$-$6031.
%\\
%Sum of components 6 and 7 in Prochaska et al. (2002b); \ion{S}{ii} and 
%\ion{Si}{ii} column densities estimated in the same way; \ion{H}{i} column density
%scaled down to match the fraction of \ion{Si}{ii} measured in the same 2 components. 
%\\ $^b$  Sum of components 3, 4 and 11 in L\'opez et al. (2002); \ion{S}{ii} and 
%\ion{Si}{ii} column densities estimated in the same way; \ion{H}{i} column density
%scaled down to match the fraction of \ion{Si}{ii} measured in the same 3 components. 
}
\end{table*}
\end{center}

%%%%%%%%%%%%%%%%%%%%%%%%%%%%%%%

\section{Introduction}

The abundance of argon in diffuse gas can be measured
from absorption line spectroscopy of   
the \ion{Ar}{i} 104.8 and 106.6 nm resonance transitions. 
Early  {\em Copernicus}  observations indicated that 
argon is moderately depleted in the local
interstellar gas  (Meyer 1989 and refs. therein), a result confirmed   by 
recent  IMAPS (Sofia \& Jenkins 1998; hereafter SJ98)
and   FUSE (Jenkins et al. 2000; Lehner et al. 2002)
observations. 
Interstellar depletions are generally attributed to the incorporation of atoms into dust grains 
and tend to be correlated   with the gas column density
(Jenkins 1987; Savage \& Sembach 1996). 
It is hard to establish whether such a correlation, suggestive of dust, exists or not for Ar
because \ion{Ar}{i} lines saturate when $N(\ion{H}{i}) > 10^{20}$ atoms cm$^{-2}$ 
 at the solar metallicity characteristic of the local interstellar gas. 
In fact, SJ98  provided  theoretical arguments to show  
that  Ar depletion is unlikely due to dust since  Ar atoms  have a low probability of 
being incorporated into dust grains.  
The Ar deficiency is most likely due to ionization effects since the  ratio between
photoionization and recombination rates is typically
one order of magnitude larger for  \ion{ Ar}{i} than for \ion{H}{i} 
(Fig. 3 in SJ98). 
Owing to this property, Lyman continuum photons 
with $h \nu > 15.76$ eV, the \ion{ Ar}{i} ionization threshold,
will be extremely efficient in ionizing \ion{Ar}{i} but not \ion{H}{i},
if they are able to leak through a \ion{H}{i} layer.  
This characteristic of Ar can be used to probe ionization conditions not only
in local interstellar clouds, 
 but   also in  clouds of external galaxies. 
In particular, with high resolution spectroscopy of QSO sources
we can study clouds in intervening galaxies % lying along the line of sight,
detected as absorption line systems.
%The absorption systems at the high end of the
%\ion{H}{i} column density distribution, known as 
Damped Ly $\alpha$ systems (DLAs), are thought to arise in these intervening galaxies 
(Wolfe et al. 1995) and offer the 
possibility to probe ionization conditions at high redshift
from the study of Ar abundances. 
% 
%These QSO absorption systems are characterized
%the QSO absorbers with $N(\ion{H}{i}) > 2 \times 10^{20}$ cm$^{-2}$
%(Molaro et al. 2001; Prochaska et al. 2002a). 
% 
% including Damped Ly $\alpha$ systems (DLAs),
%the QSO absorbers which trace high \ion{H}{i} column density gas
%at redshift $z \simeq 2/3$. 
%
At $z \gsim 2.3$
the \ion{Ar}{i} lines can in fact be detected in the optical range, 
where the collecting power of 10-m class telescopes can be exploited
to obtain
high quality QSO spectra.
Thanks to the low metallicity typical of DLAs (Pettini et al. 1999), 
the Ar resonance lines can be unsaturated
even at the high column densities, $N(\ion{H}{i}) > 2 \times 10^{20}$ cm$^{-2}$,
typical of these systems.
%A better understanding of ionization properties of DLAs
%is fundamental to probe the physics and nature of the associated galaxies
%and to accurately determine ionization corrections for
%elemental abundances studies.
% In addition, 
Therefore, DLAs   can yield information complementary to that obtained 
from studies of local ISM, where the \ion{Ar}{i}   lines become 
readily saturated.
% at the solar level of metallicity. 
Most important, if Ar abundances in DLAs are indicative of ionization conditions,
as they are in the local ISM, their study 
may yield important constraints on the nature 
and relative importance of ionizing sources at high redshift, 
a topic of particular interest in cosmology for its effects on the physical state of the
intergalactic medium and for its link to the star formation history
of the universe. % [{\bf refs..}]. 

The first detection of \ion{Ar}{i} in a DLA system was obtained 
for the $z=3.39$ absorber towards QSO 0000-26 
observed with UVES/VLT   (Molaro et al. 2001).
The Ar abundance was found
to match that of other $\alpha$-capture elements measured in the same
system, suggesting a low degree of ionization of the gas. 
Recently a few other \ion{Ar}{i} detections in DLAs have been obtained with UVES/VLT
(Levshakov et al. 2002, L\'opez et al. 2002) and   HIRES/Keck observations
(Prochaska et al. 2002a,b).
From these works there was no evidence for a general underabundance of Ar
in DLAs, with the notable exception of the $z=2.62$ system towards QSO 1759+75, 
 an absorber characterized by complex ionization conditions
(Prochaska et al. 2002a).

From the theoretical point of view,
the Ar ionization fractions in DLA systems have been estimated
by several authors, assuming different models of   radiation field
and  ionization structure of the
absorbers (Izotov et al. 2001, Vladilo et al. 2001,
Prochaska et al. 2002a). 
At variance with other metal species observed in DLAs,
the fraction of \ion{Ar}{i} is quite sensitive to the adopted model. 
This makes \ion{Ar}{i} an important  discriminant of 
ionization conditions in DLAs. Motivated by this possibility,
we have actively searched for new \ion{Ar}{i} detections in DLAs.
In this paper we present and discuss the results of this search.
The new measurements are described in Section 2, while 
in Section 3 we review the general properties of Ar abundances 
in DLAs combining these new data with previous measurements. 
We show that a significant underabundance of Ar 
is common among DLAs, contrary to previous belief, and we discuss
this result in the framework of photoionization equilibrium models.
In Section 4 we present evidence for redshift evolution of
the Ar abundances in DLAs and we briefly summarize Ar interstellar abundances
  at $z=0$. Finally, in Section 5 
we discuss these findings in the framework of current  evolution scenarios for 
DLAs and for the metagalactic radiation field. The results are summarized
in Section 6.

%                                     One column figure Fig. 1
   \begin{figure}
   \centering
 \includegraphics[clip=true,width=7.2cm]{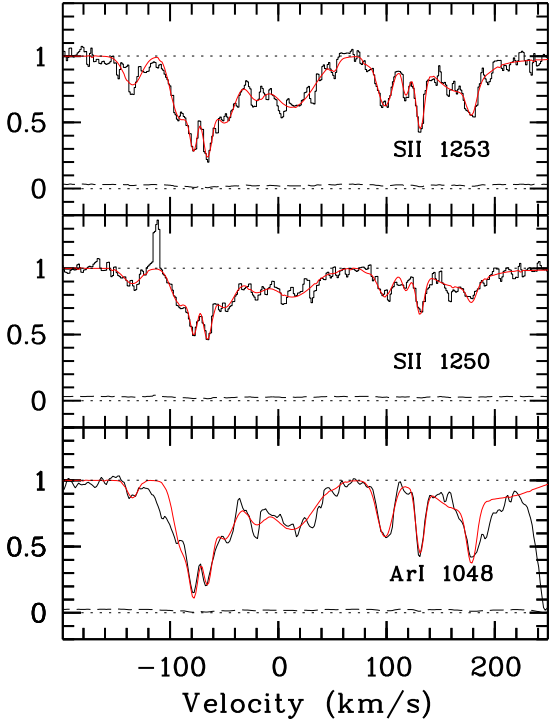} 
  \caption{Absorption profiles of \ion{Ar}{i} and \ion{S}{ii} lines
(see labels) for the DLA system at $z=2.8120$ (zero velocity)
in QSO 0528$-$2505.}
              \label{figq0528}%
    \end{figure}
%

%                                     One column figure Fig. 2
   \begin{figure}
   \centering
 \includegraphics[clip=true,width=7.2cm]{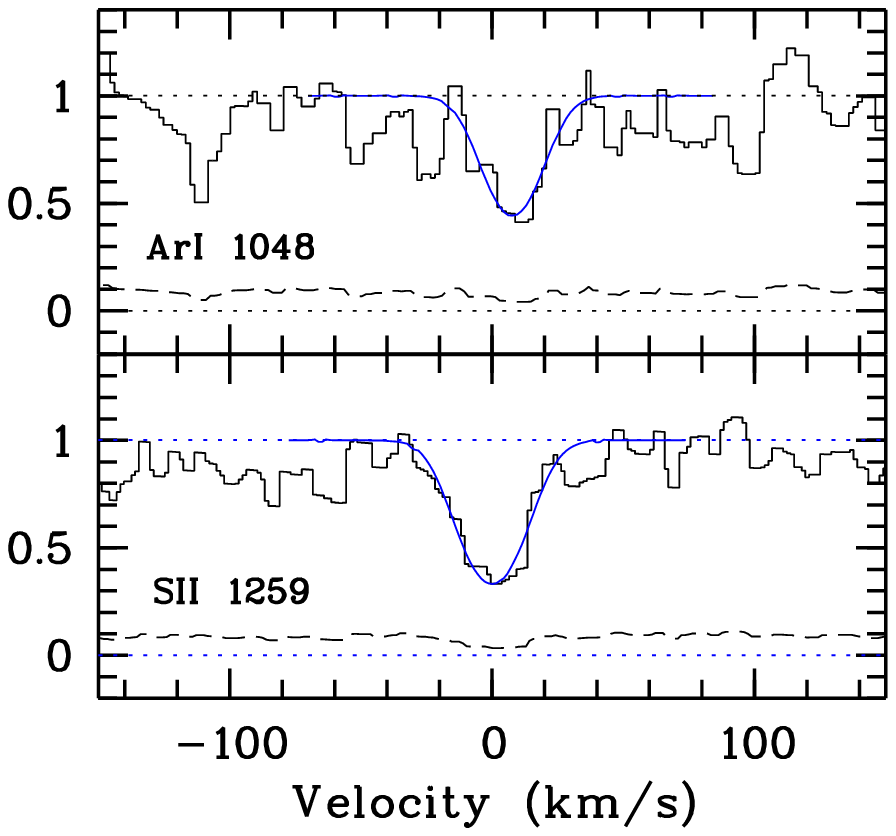} 
  \caption{ Absorption profiles of \ion{Ar}{i} and \ion{S}{ii} lines
(see labels)
for the DLA system at $z=2.3745$ (zero velocity)
in QSO 0841$+$129.}
              \label{figq0841a}%
    \end{figure}
%

%                                     One column figure Fig. 3
   \begin{figure}
   \centering
 \includegraphics[clip=true,width=7.2cm]{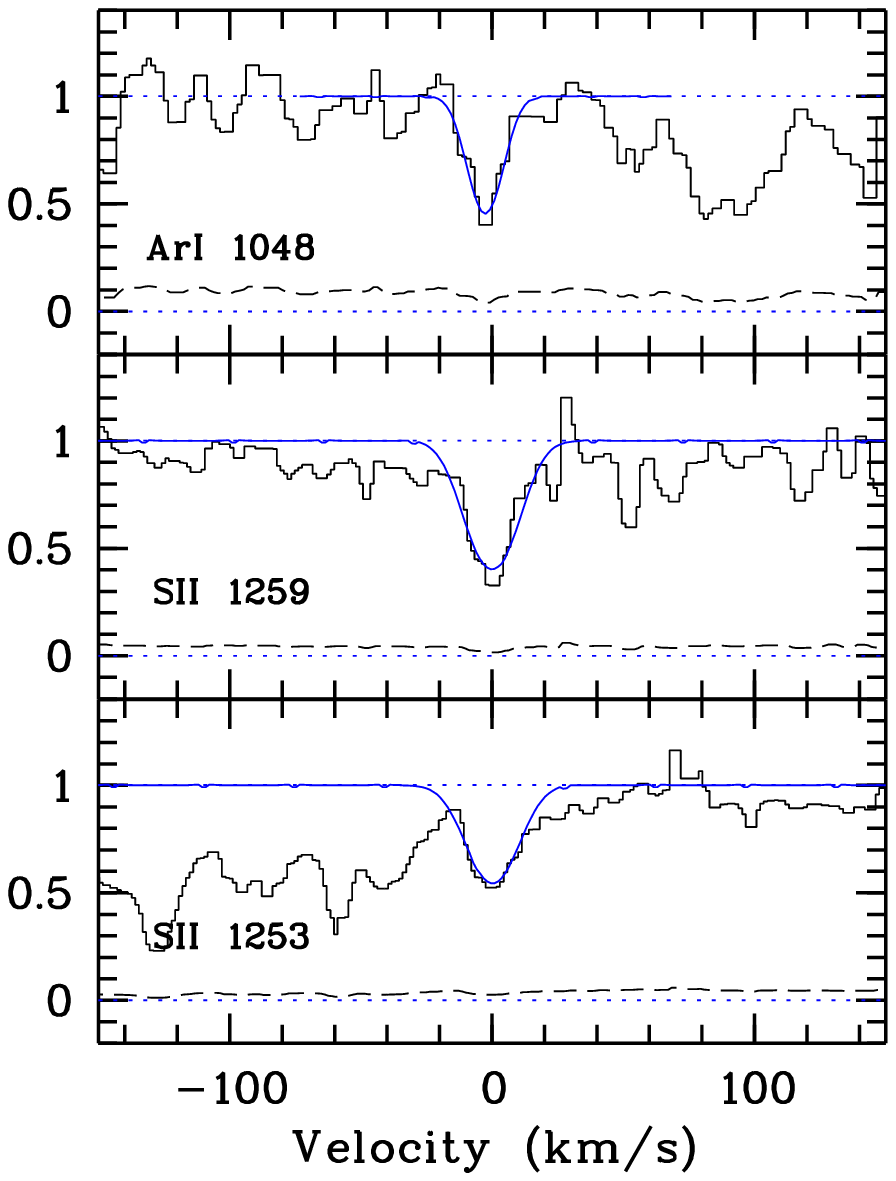} 
  \caption{Absorption profiles of \ion{Ar}{i} and \ion{S}{ii} lines
(see labels)
for the DLA system at $z=2.4762$ (zero velocity)
in QSO 0841$+$129.}
              \label{figq0841b}%
    \end{figure}
%

%                                     One column figure Fig. 4
   \begin{figure}
   \centering
 \includegraphics[clip=true,width=8cm]{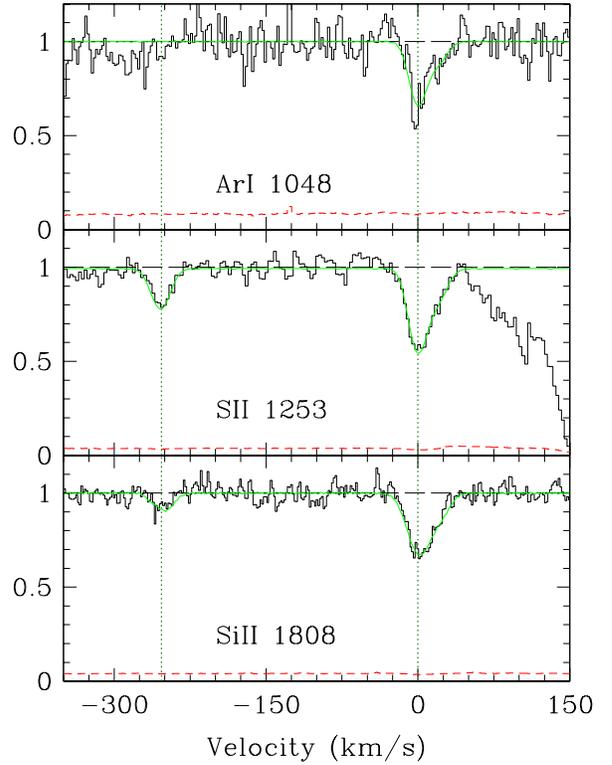} 
  \caption{ Absorption profiles of \ion{Ar}{i} and \ion{S}{ii} lines
(see labels)
for the DLA system at $z=2.4312$ (zero velocity)
in QSO 2343+1232.    }
              \label{figq2343}%
    \end{figure}

\section{The data}

The observations were obtained  with the Ultraviolet Visual Echelle Spectrograph (UVES)
fed by the Kueyen 
telescope 
(unit 2 of the ESO VLT). This spectrograph is particularly well suited for observing
the spectral region in the proximity of the ultraviolet atmospheric cutoff (Dekker et al. 2000)
allowing the \ion{Ar}{i} lines to be observed at redshifts $z \gsim 2.3$. 
The slit widths were set to 1 arcsec, yielding a mean resolving power
$R = \Delta\lambda/\lambda \simeq 4.2 \times 10^4$.  
The targets, the basic data for the DLAs investigated, and 
the new measurements of \ion{Ar}{i} column densities, $N(\ion{Ar}{i})$, 
are listed in the first 4 rows of Table 1. 
A full presentation of the data reduction  is given
in a parallel paper focussed 
on a study of nitrogen abundances in DLAs (Centuri\'on et al. 2003).
More details on UVES studies of
the systems at $z=2.812$ in QSO 0528$-$2505
and $z=2.431$ in QSO 2343+1232 can be found 
in P\'eroux et al. (2003) and D'Odorico et al. (2002), respectively.
The column densities have been derived with the Voigt-profile fitting routines FITLYMAN 
(Fontana \& Ballester 1995) implemented in the MIDAS software package. 
The oscillator strength for the transition  \ion{Ar}{i} $\lambda$104.8nm,
$f=0.2570$,  was taken
from Federman et al. (1992), the same adopted for the local ISM by SJ98
and in previous work on DLAs by Molaro et al. (2001) and Levshakov et al. (2002).
Oscillator strengths for other $\alpha$-capture elements 
were taken from   
Morton (1991), except for: \ion{Si}{ii} 1304, 1526     (Spitzer \& Fitzpatrick 
1993) and \ion{Si}{ii} 1808   (Bergeson \& Lawler 1993).

The absorption profiles of \ion{Ar}{i} lines are shown in Figs. 
\ref{figq0528}, \ref{figq0841a}, \ref{figq0841b}, and \ref{figq2343}. 
Care has been taken in assessing possible contamination
of the \ion{Ar}{i} profiles due to Ly$\alpha$ interlopers,
which would tend to increase $N$(\ion{Ar}{i}). 
Comparison with the radial velocity profiles of low-ionization species falling 
inside and outside
the Ly$\alpha$ forest has not revealed evidence of contamination. 
 In the figures we show examples of the profiles of the
\ion{S}{ii} and/or \ion{Si}{ii} lines used in the present analysis.   
The agreement with the radial velocity profiles of  \ion{Ar}{i} lines
  is  generally
excellent.
A slight misalignment between the  \ion{Ar}{i} and \ion{S}{ii}
lines is only observed  in  the $z=2.3745$ system in Q\,0841+129.
We have carefully checked that this is not due to an error in the
wavelength calibration. Some weak contamination could be responsible
for this shift, in which case the derived Ar column density could be
slightly overestimated, probably within the large error
bar of this measurement. This would not affect  the conclusion
that Ar is significantly underabundant in this system. 
Care has been taken in assessing systematic errors due to the uncertainty
of the continuum, which are largest  for the two DLAs in  Q\,0841+129
(Figs.   \ref{figq0841a} and \ref{figq0841b}).
Even in these two cases the $N$(\ion{Ar}{i}) error bars     
are sufficiently large to encompass the continuum errors.

The 4 new measurements presented here represent a significant increase
over the 6 $N$(\ion{Ar}{i}) measurements in DLAs previously obtained  
with UVES/VLT or HIRES/Keck observations, also shown in Table 1.
When also upper and lower limits of $N(\ion{Ar}{i})$ are considered
(last 5 rows of Table 1),
the combined sample sums up to a total of 15 DLAs. 
Care has been taken in comparing the abundances derived by different
authors. % and shown in Table 1.
The $f$ value of the \ion{Ar}{i} 104.8\,nm line adopted by us
is 2\% smaller than the updated Morton (1991) value adopted by 
Jenkins et al. (2000) and Lehner et al. (2002) for ISM studies
 and by Prochaska et al. (2002a,b) for DLAs studies. 
The value is 5\% larger than the old Morton (1991) value adopted by 
L\'opez et al. (2002). We did not attempt to correct for these differences,
given the presence of some saturation in the lines. 
The differences are in all cases well within the measurement errors
quoted by the authors.

\begin{figure}
   \centering
\includegraphics[width=8cm]{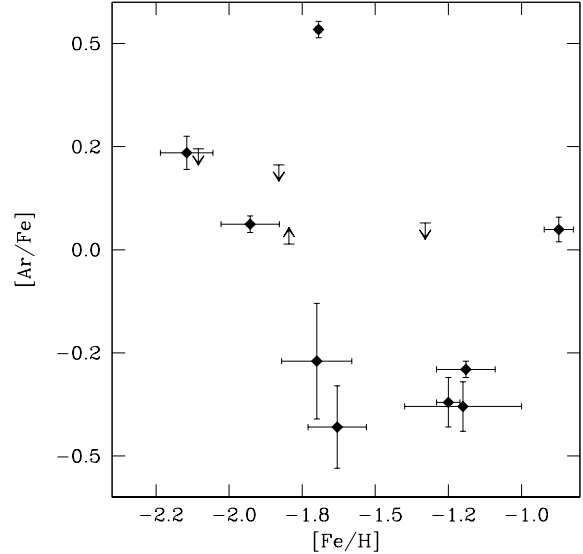} 
  \caption{[Ar/Fe] versus [Fe/H] abundance ratios in Damped Ly $\alpha$ systems. 
The sample shown in the figure includes all the systems of Table 1 for which a measurement
of $N$(\ion{Fe}{ii}) is available.
References to the original measurements are given in the table.  }
              \label{figArFe_FeH}%
    \end{figure}

For the sake of comparison with previous work,
in Fig. \ref{figArFe_FeH} we plot the [Ar/Fe] versus
[Fe/H] abundance ratios for the combined sample of DLA systems\footnote{
We follow the usual definition
$\mathrm{[X/Y]} \equiv \log \{ N(\mathrm{X}^i)/N(\mathrm{Y}^j) \} - 
\log (\mathrm{X/Y})_\mathrm{sun}$,
where $i$ and $j$ indicate the dominant ionization state in \ion{H}{i}
regions for the elements X and Y, respectively.
In this paper we adopt the meteoritic solar abundances of   
 Grevesse \& Sauval (1998), with the exception of O 
and Ar, taken from Holweger (2001) and SJ98,
respectively. }.
The  Ar/Fe ratio is likely to be affected by nucleosynthesis evolution,
dust depletion and ionization effects. Disentangling these effects is extremely
difficult, making the
interpretation of the data shown in Fig. \ref{figArFe_FeH}  quite uncertain.
As we explain below, the  ratio of Ar relative to $\alpha$-capture elements
is best suited for casting light on ionization effects alone.

%                                     One column figure Fig.5 

\section{The Ar/$\alpha$ ratio in Damped Ly $\alpha$ systems}

In columns 5, 6, and 7 of Table 1 we give the   Ar abundances
% for the combined sample,
normalized to those of O, S, and Si,
representative of $\alpha$-capture elements
measured in DLAs.  
%
%In columns 5, 6 and 7  of  Table  1 we show the resulting logarithmic
%ratios 
%of $N(\ion{Ar}{i})$ over $N(\ion{O}{i})$, $N(\ion{Si}{ii})$,
%and $N(\ion{S}{ii})$, respectively, 
%expressed in terms of solar abundances.
%
%
%This means, in practice, that the ratios shown in the table have been derived
%dividing  \ion{Ar}{i} by  
%\ion{O}{i}, \ion{S}{ii} and \ion{Si}{ii} column densities,
%when available.  
%
%From the combined sample of Table 1 we can infer general properties
%of Ar abundances in DLAs. 
Most of the systems have [Ar/O,Si,S] ratios below the solar value, 
with underabundances of about $-0.7$ dex
in several cases. 
%
%Most of the systems show depletions
%of $\approx -0.4$ to $-0.7$ dex, with 
Only the z=3.39 DLA towards Q0000-26 shows solar ratios. 
%A couple of systems show an
%intermediate behaviour, with underabundances of $\approx -0.2$ dex.
%
The [Ar/O,Si,S] ratios show therefore a significant scatter, with differences as high
as 0.8 dex. 
All upper and lower limits are consistent with this range of values. 

Because O, Si, S,  and Ar are all $\alpha$-capture elements, 
the [Ar/O], [Ar/Si], and [Ar/S] ratios
are   expected to be only weakly dependent  on the detailed
chemical history undergone by each system. 
In fact, when at least two measurements of [Ar/O,Si,S] are available
for a given absorber,
one can see in Table 1 that the ratios are in good agreement between them.
The only exception is the $z=3.025$ system in Q0347$-383$,
for which  O and Si abundances show a difference of 0.2 dex.  However,
we do  not believe that nucleosynthesis is a likely source for  explaining
the significant underabundances found in the 
[Ar/O,Si,S] ratios. 
We now consider both dust and ionization effects as possible
sources of these underabundances.

\subsection{Dust depletion}

\begin{figure}
   \centering
\includegraphics[width=8cm]{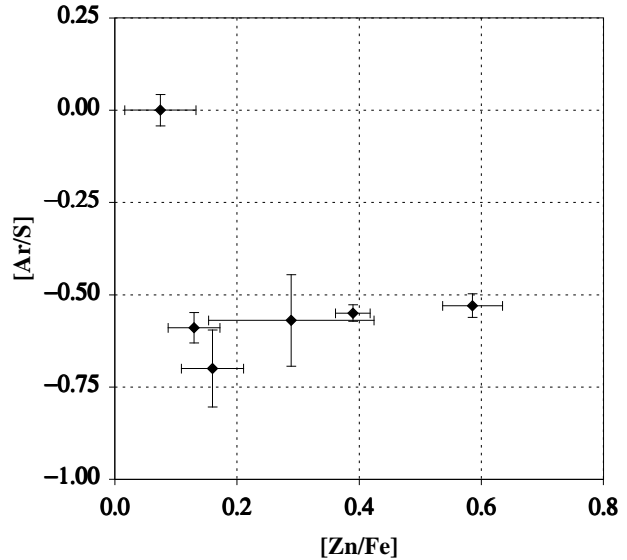} 
  \caption{[Ar/S] versus [Zn/Fe] abundance ratios in Damped Ly $\alpha$ systems. 
See Table 1 for references to the original measurements.   }
              \label{figArS_ZnFe}%
    \end{figure}

The evidence  that  Ar is unlikely to be incorporated in dust
is mostly based on theoretical arguments
(SJ98). In fact,  local interstellar  
measurements of  $N(\ion{Ar}{i})$ are feasible only 
in low-column density lines of sight 
($N(\ion{H}{i}) < 10^{20} \mathrm{cm}^{-2}$), which are not 
expected to have a significant  dust content. 
The Ar measurements in DLAs provide the possibility to test
the existence of Ar dust depletion in high redshift galaxies,
in a regime of higher \ion{H}{i} column densities.
We can estimate the dust content of individual DLAs making use of
the [Zn/Fe] ratio, given the large differential depletion between Zn and Fe,
well known from local ISM studies (Savage \& Sembach 1996) and
also supported from  studies of DLAs
(Pettini et al. 1994; Vladilo 1998; Prochaska \& Wolfe 2002)\footnote{ 
This is true even if
the nucleosynthesis of Zn is not well understood (see Prochaska \& Wolfe 2002).}.
In Fig. \ref{figArS_ZnFe} we plot
  [Ar/S]   versus   [Zn/Fe] 
for the subsample of Table 1 with
available Fe and Zn data.
Should Ar be incorporated into dust, we would expect
the [Ar/S] ratio to become more and more underabundant with increasing
[Zn/Fe], because S is known to be undepleted in the interstellar gas (Savage \& Sembach 1996). 
However, this type of trend is not seen in Fig. \ref{figArS_ZnFe}. On the
contrary, severe underabundances of Ar are present even when [Zn/Fe] is relatively
small, i.e. when the amount of dust is negligible.  
This empirical test indicates that a
mechanism different from dust depletion is responsible for the Ar 
underabundances in DLAs, lending support to the
theoretical arguments given by SJ98. 

Among the abundances reported in Table 1 only 
the [Ar/Si] ratio is possibly affected by dust depletion, given the fact that O and S
are known to be undepleted from local ISM studies. 
In the penultimate column of Table 1 we give the amount
of Si/Ar differential depletion expected for DLAs with
available Zn data, estimated following the procedure
of Vladilo (2002).

%                                     One column figure Fig. 6
%

% 
 %                                 Fig. 7                             
    \begin{figure}
    \centering
\includegraphics[width=8cm]{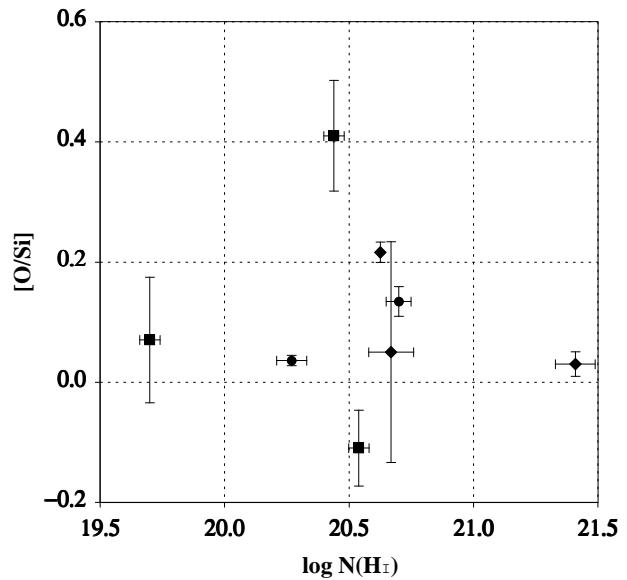} 
   \caption{The \ion{O}{i}/\ion{Si}{ii} ratio in Damped Ly $\alpha$ systems.
 %Symbols: measurements.
 Diamonds:  (from left to right): Levshakov et al. (2002),
 Dessauges-Zavadsky et al. (2001), Molaro et al. (2001). 
 Squares: Pettini et al. (2002). Circles: Prochaska et al. (2002a). 
% Lines: predictions of photoionization models
% S2 (dotted line) and H1 (dashed line)  obtained by V\&01)
% to match the observed Al$^{++}$/Al$^+$ ratios in DLAs.
% The model predictions  are calculated for a  solar
% Si/O abundance ratio.
}
               \label{figOSi_H} 
     \end{figure}
% %

\subsection{Ionization effects}

Abundance measurements in DLAs are performed 
using dominant ionization states of metals in \ion{H}{i} regions,
which can be neutral 
(e.g. \ion{N}{i}, \ion{O}{i}, and \ion{Ar}{i})
or singly ionized
(e.g. \ion{Si}{ii}, \ion{S}{ii}, \ion{Fe}{ii}, and \ion{Zn}{ii}), 
depending on the ionization potential of each element.
The predominance of such species in \ion{H}{i} regions
is well known from local ISM studies and can be easily explained 
in terms of photoionization equilibrium models of DLAs
(Howk \& Sembach 1999; Vladilo et al. 2001, hereafter V\&01;
Prochaska et al. 2002b).
Ionization corrections for the abundances can be estimated,
in the framework of these models, 
as a function of the intensity and spectrum of the 
adopted ionizing continuum.  
Unfortunately, singly ionized species can also arise in
intervening \ion{H}{ii} regions, if they
happen to lie   inside the Damped systems.
We  discuss models with and without an intervening
\ion{H}{ii} region.

\subsubsection{Intervening \ion{H}{ii} regions}

The presence of intervening \ion{H}{ii} gas in DLAs  would
yield an excess  of ionized species, but not of neutrals.  
As a consequence, they would create
an (apparent) underabundance of the Ar/Si and Ar/S ratios,
derived by comparing \ion{Ar}{i} with
 \ion{Si}{ii} or \ion{S}{ii}. On the other hand,
the Ar/O ratios, derived from neutral species, are not expected to be altered by this
effect. Therefore, if  intervening \ion{H}{ii} gas exists, we should find [Ar/Si,S] $<$ [Ar/O]. 
The only two [Ar/O] ratios in our sample  do not support this possibility
(see Table 1).    Since
this sample is too small --- and does not include cases with significant
Ar underabundances ---  
we have also investigated  the behaviour of the \ion{O}{i}/\ion{Si}{ii} ratio, 
for which  9 DLAs measurements exist,
2 of them   shown in Table 1 and the others found in the literature
(Prochaska \& Wolfe 1999, Dessauges-Zavadsky et al. 2001,
Pettini et al. 2002, Prochaska et al. 2002a). 
The resulting [O/Si] ratios are shown in Fig. \ref{figOSi_H}. 
The mean value of this sample is
$< [\mathrm{O/Si}] > = +0.10 \pm 0.14$ dex. 
A few \ion{O}{i} measurements are derived from partially saturated 130.2 nm lines,
and in these cases the real [O/Si] ratio could be even higher.
A little excess of [O/Si] may be expected  due to differential
dust depletion between the two elements (Savage \& Sembach 1996).
Only one system shows  a modest underabundance which can
be explained in the framework of the ionization balance of \ion{H}{i} regions
discussed below.   The general lack of O/Si underabundance
indicates that intervening \ion{H}{ii} gas is uncommon in DLAs.  
%\footnote{
%A little excess of [O/Si] is expected, in some cases, due to the differential
%dust depletion between the two elements.}. 
On the basis of this result, DLA ionization models which consider
 intervening \ion{H}{ii} regions
as inherent to DLAs ionization structure, such as 
the models discussed by Izotov et al. (2001), are not considered hereafter.

%                                     One column figure Fig. 8
   \begin{figure}
   \centering
\includegraphics[width=8cm]{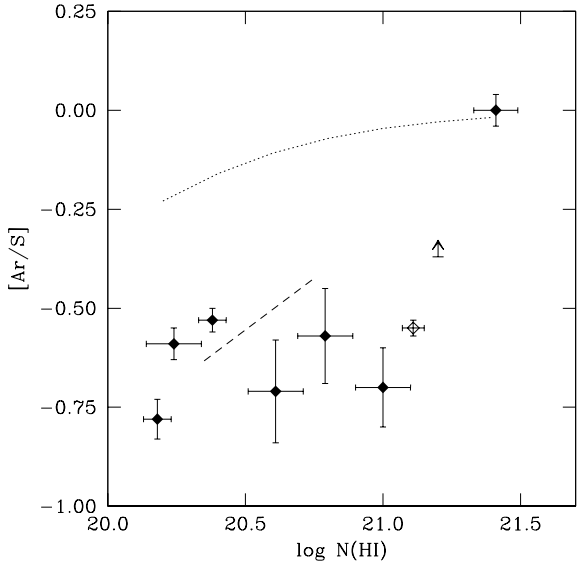} 
  \caption{The [Ar/S] ratio in Damped Ly $\alpha$ systems.
    Filled diamonds and arrow: measurements and lower limit from Table 1;
open diamond: $z=2.812$ system in Q0528$-$2505.
    Continuous curves: predictions of the photoionization models
    S2 (dotted curve) and H1 (dashed curve) described in Section 3.2.2.  
 }
              \label{figArS_H}%
    \end{figure}
%

%                                       Fig. 9      
   \begin{figure}
   \centering
 \includegraphics[width=8cm]{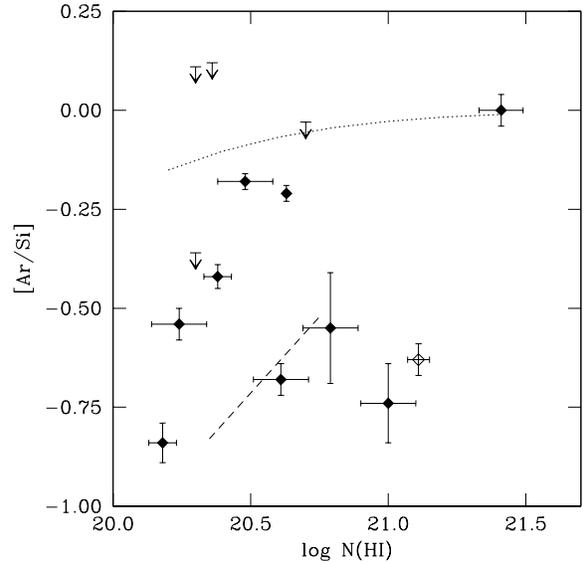} 
  \caption{The [Ar/Si] ratio in Damped Ly $\alpha$ systems.
    Filled diamonds and arrows: measurements and upper limits from Table 1;
open diamond: $z=2.812$ system in Q0528$-$2505.
Continuous curves: predictions of the photoionization models S2 (dotted curve) 
     and H1 (dashed curve) described in Section 3.2.2. 
 }
              \label{figArSi_H}%
    \end{figure}

\subsubsection{ Ionization balance of the \ion{H}{i} region }
  
Photoionization models
can be constrained when two or more ionization states of 
the same element 
  coexist and can be detected in \ion{H}{i} regions. 
This is the case of \ion{Al}{ii} and \ion{Al}{iii},
the only  species of intermediate ionization which can be easily measured in DLAs.
In fact,  \ion{Al}{iii}  
% In fact, there is sound
% observational evidence that \ion{Al}{iii} 
shares a common velocity distribution --- and, hence, a common location in the DLAs ---
with species of lower ionization 
% (e.g. \ion{O}{i}, \ion{Si}{ii} and \ion{Al}{ii}) 
 (Wolfe \& Prochaska 2000).
General models  of DLAs ionization  should be able to explain the  
abundances of \ion{Al}{iii} relative to \ion{Al}{ii} and other ions.
In a previous work,
a trend of decreasing   \ion{Al}{iii}/\ion{Al}{ii} ratio
with increasing  $N$(\ion{H}{i})   was found and 
used to  constrain
photoionization  models (V\&01). 
Two types of ionizing continuum were considered:
(1)  a hard, QSO-dominated radiation field 
(Haardt \& Madau 1996; see also Howk \& Sembach 1999)
and (2) a soft, stellar type radiation field  ($T_{eff}= 33,000$ K; Kurucz 1991). 
In the first case (H1 model)   the \ion{Al}{iii} is produced 
by the hard photons, which are able to penetrate  
\ion{H}{i} layer in depth. 
In the second case (S2 model) the  \ion{Al}{iii} originates in
a partially ionized interface created by the soft
radiation field, which is not able to penetrate the neutral gas in depth.
Both types of models are able to reproduce the
observed trend  \ion{Al}{iii}/\ion{Al}{ii} versus $N$(\ion{H}{i})
with a proper choice of the ionization parameter, $U$. 
%$U = \Phi({\rm H})/cn_{\rm H}$, where  $\Phi({\rm H})$ is the total surface flux of
%ionizing photons (cm$^{-2}$ s$^{-1}$) and $n_{\rm H}$ is the hydrogen particle
%density (cm$^{-3}$). 
%
For model H1, the requirement to fit the observed
\ion{Al}{iii}/\ion{Al}{ii} versus $N$(\ion{H}{i}) trend implies that
the ionization parameter must vary as $U \propto N(\ion{H}{i})^{-1.5}$.
%with representative values $\log U = -4.2$ and $-4.8$ at 
%$\log N(\ion{H}{i}) = 20.3?$ and $20.75$, respectively. 
However, the model H1 cannot be computed at arbitrarily high values
of $N$(\ion{H}{i}), owing to uncertainty in the thermal
solution of the photoionization computations.
For model S2 the trend \ion{Al}{iii}/\ion{Al}{ii} versus $N$(\ion{H}{i}) 
can be reproduced with a constant ionization parameter.
%with typical values $\log U = -2.2_{-0.4}^{+0.5}$. 
We refer to V\&01 for more details. 
After constraining the ionization parameter in such a way  to match the \ion{Al}{iii}/\ion{Al}{ii} trend,
the H1 and S2 models can be used to predict
the behaviour of Ar/$\alpha$
ionic ratios  as a function of $N$(\ion{H}{i}).
In Figs. \ref{figArS_H} and \ref{figArSi_H} we show the  
predictions for the
\ion{Ar}{i}/\ion{S}{ii} and  \ion{Ar}{i}/\ion{Si}{ii} ratios, respectively,
derived with model H1 (dashed line) and S2 (dotted line),
for a gas with solar abundance ratios. 
Both models predict that
deviations from the solar value induced by ionization become
less important  with increasing $N$(\ion{H}{i}). 
The comparison of   observations with   model predictions 
in Figs. \ref{figArS_H} and \ref{figArSi_H} yields the following results.

1. The   [Ar/S] and  [Ar/Si] ratios, when available for a given DLA, 
give consistent indications
on the ionization conditions\footnote{
The individual DLAs in the figures can be recognized by means
of their \ion{H}{i} column density, given in Table 1.
}. 
For instance, both ratios
of the $z=3.390$ system in QSO 0000-26
are in agreement with model S2.  
For the $z=2.462$ system in QSO 0201+365
the ratios  yield consistent  results,
slightly above the H1 line,
when the Ar/Si dust correction term is taken into account
(Table 1; dust corrections are negligible in the other cases). 
The concordant indications obtained from different ratios, each one measured
and modelled independently, indicates that the results are,
at least, self consistent. 
 
2. Only a few [Ar/S,Si] ratios   lie in the proximity of the S2 curve,
suggestive of a soft ionizing continuum. 
Most of the data lie instead below the S2 model, with several cases
close to the H1 curve, suggestive of   a hard continuum.

3. Even if the DLA with highest $N$(\ion{H}{i}) in our sample 
(the $z=3.390$ system in QSO 0000-26) does
show negligible ionization effects, 
there is no evidence for the systematic decrease of ionization
effects with increasing
$N$(\ion{H}{i}) predicted by the models. 
The trend  could be   smeared out   because    
different DLAs are embedded in different types of
ionizing continuum,
with  one possible case of extremely intense  continuum,
as we mention in the next point. 

4.  The $z=2.812$ system in QSO 0528-25   is   peculiar 
 since its redshift is larger than the emission redshift
of the QSO, $z_\mathrm{em} = 2.779$. 
Most likely the system has a large, positive radial velocity 
and is very close to the QSO. In this case, 
the large [Ar/$\alpha$] underabundance at high $N(\ion{H}{i})$
observed  in this system
(open symbol in Figs.    \ref{figArS_H} and \ref{figArSi_H})
could be due to the effect of a strong, QSO-type 
radiation field enhanced relative to the 
diffuse background.  A detailed analysis of this DLA will be presented in  a
separate paper (P\'eroux et al. 2003).

%                                   Fig.10
%                                     
   \begin{figure}
   \centering
 \includegraphics[width=8cm]{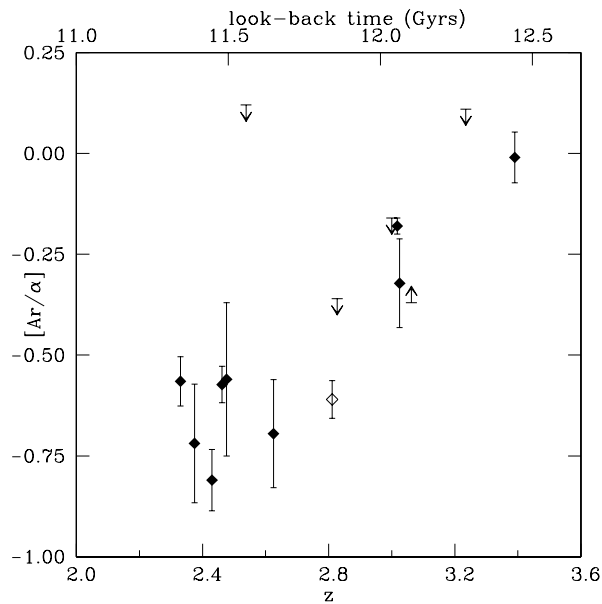} 
  \caption{The [Ar/$\alpha$] ratios in DLAs plotted
versus   redshift (bottom axis) and  
look-back time (top axis), for
$H_0=65$ km s$^{-1}$ Mpc$^{-1}$, $\Omega_m=0.3$, and $\Omega_\Lambda=0.7$. 
See Table 1 and Section 4.  }
              \label{figAr_z}%
    \end{figure}

\section{Redshift evolution of Ar abundances }

We now present evidence for redshift evolution
of Ar abundances, based on measurements in DLAs 
in the range $2.3 \leq z \leq 3.4$ and on local ISM
measurements,  representative of the conditions at $z=0$. 

\subsection{Ar abundances in DLAs} %
%at $2.3 \leq z \leq 3.4$}

In Fig. \ref{figAr_z} we show   
%[Ar/$\alpha$]   versus $z$   to search for a possible
%redshift dependence. The results   shown  in the figure
% represents a synoptic view of 
all [Ar/$\alpha$] measurements
and limits given in Table 1 plotted  versus $z$. 
For the limits, we adopt the most stringent limit on [Ar/$\alpha$] available.
For the measurements, 
we define [Ar/$\alpha$] 
as the mean value of available [Ar/O], [Ar/S], and [Ar/Si] data, with error bars
obtained from error propagation of individual data. 
Only for the $z=3.025$ absorber in QSO 0347-38 the individual error bars
are smaller than the difference between the [Ar/O] and [Ar/Si] measurements.
In this case 
we adopt an error bar sufficiently large to encompass both
measurements. 

The [Ar/$\alpha$] ratios in Fig. \ref{figAr_z} 
increase with redshift, from $\sim -0.7$ dex at $z \simeq 2.4$
to $\simeq 0$ dex at $z \simeq 3.4$.
The trend is supported by all the 10 measurements and the 5 limits,
four of which happen to be stringent. 
A linear regression   of   [Ar/$\alpha$]  
versus  $z$ for the 10 measurements
yields a slope $m=0.63 \pm 0.12$ at 99.92\% confidence level (cl). 
If the $z=2.812$ system in Q 0528$-25$ is excluded, 
the linear regression yields $m=0.65 \pm 0.11$ at 99.94\% cl. 
The trend versus $z$ is particularly remarkable given the lack of
any trend  versus $N(\ion{H}{i})$ as discussed above. 
This result strongly suggests that Ar abundances are subject to evolution,
even if the sample is still small at $z \gsim 3$   
to allow firm conclusions to be drawn. In favour of the
existence of the trend, we note that
the linear regression analysis still yields consistent results, i.e.
 $m=0.62 \pm 0.17$ at 98.91\% cl, 
after excluding the data point at $z=3.4$   (8 measurements;
$z=2.812$ system also excluded).

To our knowledge, the Ar/$\alpha$ ratio shows the strongest redshift variation
of relative  abundances  found so far in DLAs. 
Ionic ratios predicted to be
less affected by the ionization field do not show such variations.
For instance, the Si/Fe ratio shows a nearly
constant value, although with significant scatter,
 over the large redshift interval $1.8 \lsim z \lsim 4.2$
(Fig. 26 in Prochaska \& Wolfe 2002). Once corrected for
dust effects, the Si/Fe ratio does show some increase
with redshift (Fig. 3 in Vladilo 2002) which, however, can be explained
in term of  chemical evolution
(Fig. 5 in Calura, Matteucci \& Vladilo 2003).   
As mentioned before, we do not expect 
significant chemical evolution for the Ar/$\alpha$ ratio. 
The interpretation of the observed evolution
of Ar/$\alpha$  versus redshift is discussed in Section 5.

\subsection{Interstellar Ar abundances at $z=0$}

Interstellar lines of sight with $\log N(\ion{H}{i}) \gsim 20.3$ 
through the Milky Way or nearby galaxies  
are representative of the behaviour of DLAs at $z=0$. 
Most interstellar Ar abundances have been obtained 
for the Milky Way. 
Unfortunately, at the solar metallicity typical of the Milky Way,
  \ion{Ar}{i} interstellar  absorptions  are saturated
when  $\log N(\ion{H}{i}) \gsim 20.3$ and
reliable measurements can be obtained 
only in lines of sight 
with column densities somewhat lower than in DLAs. 
This difference is taken into account in the discussion below (Section 5).  

Milky-Way measurements of [Ar/$\alpha$] are available 
in   4 lines of sight, 
yielding  $-0.3 \leq$ [Ar/O] $\leq 0.0$ dex  (Jenkins et al. 2000; Lehner et al. 2000).
Since the local ISM has solar metallicity, the Ar underabundance
can also be estimated from the [Ar/H] ratios, which are available for 3 other lines
of sight, yielding [Ar/H] = $-0.6$, $-0.4$, and $-0.2$ dex (SJ98). 
Limits obtained from saturated  \ion{Ar}{i} lines are consistent with
these values.  
The median value of all available measurements is $-0.2$ dex,
significantly larger than the typical [Ar/$\alpha$] ratio
in DLAs at $z\sim 2.5$. 

The largest Ar deficiencies are found in the Galactic plane. Only one,
observed towards $\beta$ Cen  ([Ar/H] $=-0.6$ dex),
is comparable to those measured in DLAs at $z \sim 2.5$. 
According to SJ98, this large deficiency
requires the cloud to be located close to an ionizing star
or to be embedded in soft X-ray radiation produced by local hot gas. 
The existence of soft X-ray emitting gas in hot
cavities of the local \ion{H}{i} distribution is well known
and the line of sight to $\beta$ Cen
intersects the Local   Bubble (Kuntz \& Snowden 2000; Sfeir et al. 1999)
and, probably, a further hot cavity associated with Loop I (Iwan 1980; Centuri\'on \& Vladilo 1991). 
Therefore, the severe Ar deficiency measured towards $\beta$ Cen may    
well represent a particular, rather than general, interstellar condition.

Among the local ISM measurements, those taken at high Galactic latitude
have better chance to reflect the influence of radiation fields external to the Milky Way.
Only three lines of sight have been investigated at $|b| > 35^\circ$,
yielding [Ar/O] = $-0.17$, $-0.18$, and $-0.27$ dex (Jenkins et al. 2000; Lehner et al. 2000),
consistent with the median value of the full sample of Milky-Way
measurements, $-0.2$ dex.

To our knowledge, the only \ion{Ar}{i} interstellar absorption detected in a nearby
galaxy is the one obtained by Vidal-Madjar et al. (2000)  from a FUSE spectrum of
 IZw18 taken with a large entrance aperture covering the whole galaxy.
Analysis of the absorption spectrum yields [Ar/Si] $\simeq -0.3$ dex (Izotov et al. 2001),
even though a value somewhat closer to solar can be derived by taking into account
the effects of large-scale velocity fields integrated in the field of view  (Levshakov et al. 2001). 

In summary, with the available data set
it seems reasonable to take [Ar/$\alpha$] $\sim -0.2$ dex
 as a representative  interstellar value of the local universe.
The comparison with  the low values observed 
in DLAs  at $z \sim 2.5$  indicates that the Ar abundance must significantly increase  
from  $z \sim 2.5$ to $z=0$.

\section{Discussion}

The present study indicates   that the full range of
[Ar/$\alpha$] values observed in DLAs 
can be reproduced in the framework of photoionization models
of \ion{H}{i} regions by properly tuning the spectrum of the ionizing continuum.
Even if the models considered are quite idealized, it  is clear that the
magnitude of the underabundance is related to the shape of
the ionizing spectrum,  [Ar/$\alpha$] $\lsim -0.5$ dex being suggestive 
of a hard, QSO-type continuum, 
while  [Ar/$\alpha$] closer to solar value being suggestive of a soft, 
stellar-type continuum\footnote{Strictly speaking, the latter indication 
must be taken with caution at high column densities
since we are not able to compute
the ionization balance for model H1
when $\log N(\ion{H}{i}) \gsim 21$ (see 3.2.2).}.
The unique capability of \ion{Ar}{i}
to discriminate between different sources of ionization 
at work in DLAs offers the opportunity
to cast light on the origin of the ionizing continuum in the associated galaxies,
which in principle will include both internal contributions,
due to stellar emission,
and external ones, due to integrated metagalactic radiation. 
The  origin  of the integrated UV background has
been the object of intensive investigation, with
QSOs and star-forming galaxies being
considered as important contributors to the UV background
at various redshifts
(Haardt \& Madau, 1996; %Songaila \& Cowie 1996;
%Savaglio et al. 1997; 
Songaila 1998; Steidel et al. 2001; Bianchi et al. 2001; Haehnelt et al. 2001). 
From these studies a picture is emerging
in which the intensity and spectral shape of
the UV background evolves with redshift.  
These studies are necessarily linked to those of the global
star formation rate in the universe,
which   also shows evidence for evolution 
(Madau, Pozzetti, \& Dickinson 1998; Lanzetta et al. 2002).   
Both types of change (UV background and star formation rate) 
are expected to affect the ionization state 
and the Ar abundances in DLAs. 
%On the basis of our previous discussion,
The trend   of Fig. \ref{figAr_z}
suggests that the ionizing continuum in DLAs becomes  gradually harder   
from $z \sim 3.4$ up to $z \sim 2.3$. The interstellar Ar abundances
in the local universe suggest that the continuum should become softer 
from $z \sim 2.3$ to $z=0$. 
We   discuss the possible  implications of such evolutionary trends. 

\subsection{Ionization properties between $z \sim 3.4$ and  $z \sim 2.3$ }
 
The general underabundance of Ar  at $z \sim 2.5$ 
is quite clearly established with the present data set. This result suggests that 
the spectrum of the continuum is relatively hard, a conclusion
which in turn supports an external origin of DLAs ionization,
with dominant contribution from QSOs around  redshift   $z \sim 2.5$. 
This interpretation is consistent with
the   high space density of quasars observed in the redshift interval
  $1.5 \lsim z \lsim 3$
(Shaver et al. 1996). 

The interpretation of the
modest or negligible underabundances seen at $z \gsim 3$,
suggestive of   a softer ionizing spectrum, 
is less straightforward.
The decline of space density of quasars for $z>3$  (Shaver et al. 1996)
may lead to a softening of the  UV continuum  
if quasars are gradually replaced by stars as dominant sources of ionization.
However, 
we do not know whether the bulk of the soft photons required 
is internal (e.g.  stars inside the DLAs)
 or external 
(e.g. Lyman-break galaxies).
%as suggested by results from Steidel et al. 2001). 
We briefly discuss  these two possible scenarios, taking also into account
the possibility that the softening of the ionizing continuum at $z>3$ may
reflect an evolution of the optical depth of the IGM.

{\em UV continuum at $z > 3$ dominated by internal stars.}
An internal origin of the flux
requires a strong synchronization of stellar emissivity 
among DLAs, in such a way that internal stellar photons
dominate at $z>3$ and become negligible at  $z<3$ in most cases.
This requirement 
is rather stringent since it would imply a strong decrease of the star formation rate
in DLAs at $z \sim 3$, not easy to understand given
the weak evidence for redshift evolution of metallicities
(Pettini et al. 1999, Prochaska \& Wolfe 2000, Vladilo et al. 2000)
and relative  abundances (Vladilo 2002)  in DLAs.
In addition, high redshift DLAs are likely to have low dust content and this
makes more critical the requirement that
the UV continuum is dominated by internal starlight. In fact,
we know from Milky Way studies that the main contribution to
the diffuse interstellar radiation is given by photons
scattered by dust (see Bowyer 2001).
If the dust content is low, the UV photons would tend
to escape from the system rather than ionize the \ion{H}{i}
region(s) inside the system.

{\em UV continuum at $z > 3$ dominated by external galaxies.}
In this scenario 
the external continuum   dominates the internal one
in the whole redshift interval explored,
with a   spectral distribution  gradually changing from hard, QSO-type at
$z < 3$ to  soft, stellar-type at $z > 3$. 
This  requires
(i) the internal emissivity from stars to be generally modest,
%(ii) the  \ion{H}{i} regions in DLAs to be directly exposed to the external continuum, and
(ii) the UV background must become softer at $z>3$.   
These requirements 
seem easy to accommodate within our current views of DLAs,
LBGs and studies of metagalactic UV radiation.
  In fact, the modest stellar emissivity of DLAs
is consistent with an origin in galaxies   characterized by modest star formation rates,
as suggested by studies of chemical evolution     
of DLAs  (Calura et al.  2003).
A low stellar emissivity  is also suggested by searches for
Ly\,$\alpha$ emissions 
(see  Fynbo et al. 1999; Fynbo et al. 2000)
and H\,$\alpha$ emissions from DLAs
(Teplitz, Malkan \& McLean 1998; Bunker et al. 1999; Bouch\'e et al. 2001)
which have led, in general, to negative or modest
detections. 
A softening of the spectrum of the metagalactic
UV continuum above $z \simeq 3$  
is supported by  studies of \ion{Si}{iv}/\ion{C}{iv}
ratios in metal absorption systems (Savaglio et al. 1997; Songaila 1998; see
however Kim, Cristiani \& D'Odorico 2002; Levshakov et al. 2002b). 
%In fact,  
%ionizing sources alternative to QSOs  must exist at high $z$ to explain
%the high degree of ionization of the IGM in spite of the
%decline of space density of quasars for $z>3$  (Shaver et al. 1996). 
If the escape fraction of UV photons from  Lyman-break galaxies (LBGs)
at $z>3$ is high, as  suggested by Steidel et al. (2001), then the change
of the spectrum could be due to the rise of the contribution from LBGs  
parallel to the decline of the  space density of quasars. 
Perhaps more interesting is the possibility that the softening of the
continuum is related to a change of the optical depth of the IGM,
such as the one expected during the \ion{He}{ii} reionization epoch.
After this reionization, photons with $h\nu \geq 54$ eV are free to propagate
through the IGM and can
easily affect the relative abundance of \ion{Ar}{i}. In fact,      
  the ratio of the ionization over recombination rates
  is one order of magnitude higher for \ion{Ar}{i} than for \ion{H}{i}
 at $h\nu \gsim 60$ eV
(Fig. 3 in SJ98). 
Evidence for the detection of the  \ion{He}{ii} reionization has been recently reported by
Bernardi et al. (2002) and Theuns et al. (2002). 
The coincidence between the redshift interval in which the Ar abundances change
(Fig.  \ref{figAr_z}) and the interval indicated by Theuns et al. (2002)
for the  \ion{He}{ii} reionization to occur ($ 3.0 \lsim z \lsim 3.4$)
is indeed quite remarkable.  
%i.e. of ionization conditions, seems to occur on a time scale of $\lsim 1$ Gyr. 
%
In any case it is clear that, from a purely observational point of view,
the cumulative evidence for a change of the physical conditions of the IGM
around $z \sim 3$ starts to be compelling.

\subsubsection{ Optical depth of the \ion{C}{iv}/\ion{Si}{iv} layers in DLAs } 

The hypothesis that Ar abundances are governed
by radiation fields external to DLAs implies that 
the \ion{H}{i} regions   must be directly exposed to such radiation.
This bears implication on the optical depth of the 
\ion{C}{iv}/\ion{Si}{iv} layers in DLAs, which are
kinematically  disconnected from the layers of low ionization
(Wolfe \& Prochaska 2000) and probably envelope  the low ionization gas,
in a way similar to what observed in the Milky Way, where 
\ion{C}{iv} and \ion{Si}{iv} layers   are stratified at large distances from
the Galactic disk. In fact,
in order to prevent absorption from external radiation,
the high-ionization layers in DLAs should have a negligible
Lyman discontinuity. This in turn implies that the neutral
hydrogen column density associated with the highly
ionized gas should be  $N(\ion{H}{i}) \lsim 10^{17}$ atoms cm$^{-2}$.
 
\subsubsection{Soft-X origin of Ar underabundances in DLAs}

The local ISM studies suggest that 
strong Ar deficiencies might be expected in DLAs
if regions of hot, soft X-ray emitting gas exist in
proximity of the \ion{H}{i} region sampled by the line of sight.
In principle, this possibility   could be used to
interpret  the trend seen in Fig. \ref{figAr_z} as entirely due to
internal sources dominating at all redshifts (stars at $z>3$ and
soft-X bubbles at $z\sim 2.5$). However, in addition to the
drawbacks of the ``internal scenario" mentioned in Section 5.1, 
this would also require a very large filling factor of hot gas in DLAs at $2.4 < z < 3$. 
Since in the Milky Way we have detected only
one strong Ar deficiency of internal origin out of seven measurements,
such cases do not need to be frequent in DLAs.

\subsection{Ionization properties between $z \sim 2.3$ and $z =0$}

The local ISM ionization, representative of the
conditions at $z=0$, is governed
by the total contribution of ionizing sources internal and external 
to the Milky Way.
The origin of the  external  background in the far UV band
is not well understood, even though it is clear that 
(i) some external contribution does exist at high Galactic latitude,
(ii)  sources with soft continuum,
such as Galactic starlight scattered by dust and  integrated light from galaxies,
 do give some contribution;
(iii) sources with hard continuum, such as
 integrated QSOs emission and \ion{He}{ii} Ly $\alpha$ emission from the IGM,
are not   dominant  
(Paresce \& Jakobsen 1980; Bowyer 2001; Henry 2002). 
At low Galactic latitude most of the internal background 
is due to starlight scattered by dust. 
Therefore the ionizing continuum in the ISM
at either low and high latitudes
is expected to be soft and, as a consequence, the local Ar deficiencies to be small.
The modest  underabundances of Ar found in the local ISM, 
with median value $-0.2$  dex, are
consistent with this expectation. 
The fact that local Ar abundances are measured in lines of sight
with \ion{H}{i} column density lower than in DLAs
does not affect this result.   
In fact,  these lines of sight are less self-shielded 
and should be more affected by ionization.
A hard continuum would   create in any case
severe Ar underabundances, which are instead uncommon.   

If we take the   Ar measurements in the
local ISM as representative of the universe at $z=0$, 
the present results indicate that the Ar abundances in DLAs
must increase from $\sim -0.7$ dex at $z \sim 2.4$ to $\sim -0.2$ dex
at $z=0$. 
This predicted increase  is consistent
with the evolution of the relative contribution of QSOs
and galaxies to the UV background in the same redshift interval, derived from   
number density studies of the Lyman $\alpha$ forest.
According to these studies 
%at low redshifts 
the contribution of galaxies overcomes that of QSOs after $z \sim 1$  
(Bianchi et al. 2001). 
If this is true, we predict   to detect a 
rise of Ar abundances  at similar, intermediate redshifts.

\section{Conclusions}

From the  analysis of the combined sample of 
Ar abundance, obtained from  \ion{Ar}{i} absorptions in DLAs 
(10 measurements and 5 limits),
we have derived the following results. 

Most DLAs show  significant Ar underabundances
relative to other $\alpha$ elements (O, Si, and S), with
  deficiencies     as low as $\simeq -0.8$\,dex in some cases.

Departures from cosmic abundances of this magnitude are not expected 
for elements that share a common  nucleosynthetic history.
We have considered dust and ionization effects as
possible explanations of the observed underabundances.

From a study of    [Ar/$\alpha$] versus [Zn/Fe] ratios in DLAs
we have shown that dust depletion does not give a viable explanation.
This result is consistent with the lack of
dust depletion  expected for Ar 
from theoretical studies of the local interstellar medium.  

As far as ionization is concerned, we have first considered    
the possible effects of intervening \ion{H}{ii} regions inside the DLAs
and, on the basis of a study of  \ion{O}{i}/\ion{Si}{ii} ratios,
we have concluded that such  \ion{H}{ii} regions must be uncommon %in DLAs
and cannot explain the observed underabundances.
 
The full range of the observed Ar abundances can instead be
reproduced by photoionization models of   \ion{H}{i} regions embedded in
an ionizing continuum with variable spectral distribution.
By using models tuned to match  the  \ion{Al}{iii}/\ion{Al}{ii} ratios 
measured in DLAs we find that the largest deficiencies of Ar can be explained
by a hard, QSO-dominated continuum, while the modest deficiencies
by a soft, stellar-type continuum.  

We have found evidence for
a redshift dependence of the Ar abundances in DLAs.
At $z \sim 2.5$, where most of the
measurements are concentrated, the deficiencies are strong 
($-0.8/-0.6$ dex).
At $z\gsim3$  
the deficiencies are smaller ($> -0.5$ dex), with the Ar abundance
becoming solar at $z \sim 3.4$.
More measurements at $z\gsim3$ are required to understand
how general this latter result is.

The strong Ar deficiencies at $z \sim 2.5$
 indicate that the ionization is dominated
by a QSO-type spectrum, which we associate with a 
QSO-dominated metagalactic background.
The modest deficiencies  measured at higher
redshifts   suggest a predominance
of a soft, stellar-type spectrum at $z>3$.
We have considered an origin of this soft spectrum   
in both starlight  internal to DLAs and
galactic emission external to DLAs.
In both cases, the gradual change of the ionizing continuum  from
$z \simeq 2.4$ to   $z\simeq 3.4$  
poses strong requirements on the nature of DLAs and/or   the
origin of the metagalactic background. 

If the redshift variation of Ar abundances is  due to the evolution of internal, stellar emission, 
it requires a synchronization of evolution of DLAs, with most systems having strong
star formation rates at $z > 3$ and weak at $z < 3$. This seems to be
at odd with the lack of clear signal of evolution of DLAs abundances. 
Also the need for dust at very high redshift, required to maintain an internal,
diffuse source of UV radiation, seems to be difficult to reconcile with
the low dust content expected in the early evolutionary stages of DLAs.

If the trend is induced by the evolution of the metagalactic ionizing
background, it requires (i) the \ion{H}{i} regions in DLAs
to be directly exposed to the external background,
(ii) the contribution from
internal emission to be modest, and
(iii) the external background to become softer at $z>3$. 
The first requirement would imply that the \ion{Si}{iv}/\ion{C}{iv} layers
associated to DLAs should be optically thin to ionizing radiation.
The second requirement is  
consistent with previous  claims of low star-formation rates 
  based on abundance studies and
  searches for Ly\,$\alpha$ and H\,$\alpha$ emission in DLAs.  
The third requirement is consistent
with the evolution of the metagalactic spectrum inferred from 
 \ion{Si}{iv}/\ion{C}{iv} measurements in the IGM,
with claims   that LBGs at $z>3$ may have
 a high escape fraction of ionizing
photons, and with the recent finding that
the \ion{He}{ii} reionization does occur between $z \sim 3.4$ and $z \sim 3.0$.

We have summarized the results of studies of
Ar abundances in the local ISM taken as representative of a DLA system at $z=0$.
The typical underabundance of  Ar is $\sim -0.2$ dex, consistent with the expectation that the ionization
in the local universe is dominated by sources
with soft continuum. Only in one case a relatively
strong underabundance of Ar is found, which may originate
in gas embedded in hot, soft-X ray emitting gas known to
exist in the local ISM. 
Taken together, the measurements of Ar abundances in DLAs
and in the local universe appear
to offer a new tool for probing the redshift evolution of the
ionization conditions in the universe. 
Present observational limitations prevent us to use
this tool  to probe intermediate redshifts for   
determining the epoch at which QSOs  stop being dominant
contributors  to the metagalactic UV background. This epoch is 
 estimated to be at $z \sim 1$ from recent studies of the    
  Ly $\alpha$ forest. An independent assessment
based on Ar abundances will be possible from
observations with  UV spectrographs fed by
space-born telescopes with large collecting areas      
filling the gap of \ion{Ar}{i} measurements between $z=0$ and $z \simeq 2.3$.

\begin{acknowledgements}
We have benefitted from useful discussions with P. Molaro. 
We thank the referee for  suggestions that have
improved the presentation of this work.
CP is supported by a Marie Curie fellowship. 
\end{acknowledgements}

\end{document}